\documentclass[aps,preprint,pra,showpacs]{revtex4}
\usepackage{graphicx}
\usepackage[pdftex]{color}

\begin{document}

\title{Quantum to Classical Transitions via \\Weak Measurements and Post-Selection}

\author{Eliahu Cohen}
\affiliation{H.H. Wills Physics Laboratory, University of Bristol, Tyndall Avenue, Bristol, BS8 1TL, U.K}

\author{Yakir Aharonov}
\affiliation{School of Physics and Astronomy, Tel Aviv University, Tel Aviv 6997801, Israel\\ and \\Schmid College of Science, Chapman University, Orange, CA 92866, USA\\}

\date{\today}

\begin{abstract}
Alongside its immense empirical success, the quantum mechanical account of physical systems imposes a myriad of divergences from our thoroughly ingrained classical ways of thinking. These divergences, while striking, would have been acceptable if only a continuous transition to the classical domain was at hand. Strangely, this is not quite the case. The difficulties involved in reconciling the quantum with the classical have given rise to different interpretations, each with its own shortcomings. Traditionally, the two domains are sewed together by invoking an ad hoc theory of measurement, which has been incorporated in the axiomatic foundations of quantum theory
This work will incorporate a few related tools for addressing the above conceptual difficulties: deterministic operators, weak measurements and post-selection. Weak Measurement, based on a very weak von Neumann coupling, is a unique kind of quantum measurement with numerous theoretical and practical applications. In contrast to other measurement techniques, it allows to gather a small amount of information regarding the quantum system, with only a negligible probability of collapsing it. A single weak measurement yields an almost random outcome, but when performed repeatedly over a large ensemble, the averaged outcome becomes increasingly robust and accurate. Importantly, a long sequence of weak measurements can be thought of as a single projective measurement.
We claim in this work that classical variables appearing in the macro-world, such as centre of mass, moment of inertia, pressure and average forces, result from a multitude of quantum weak measurements performed in the micro-world. Here again, the quantum outcomes are highly uncertain, but the law of large numbers obliges their convergence to the definite quantities we know from our everyday lives.
By augmenting this description with a final boundary condition and employing the notion of ``classical robustness under time-reversal'' we will draw a quantitative borderline between the classical and quantum regimes. We will conclude by analyzing the role of macroscopic systems in amplifying and recording quantum outcomes.
\end{abstract}


\maketitle

\section{Introduction}
\label{ECintro}

The vague border between the classical and quantum realms gives rise to the well-known measurement problem. The problem is best understood by considering the unique properties of the quantum state space, which is boosted in size compared to the classical phase space, in order to accommodate distinctly non-classical entangled states and states of superposition. The former entails nonlocal correlations; the latter, stemming from the mutual incompatibility of conjugate observables, implies that the quantum reality cannot be accounted for in classical terms of definite physical properties. \\

However, a superposition is never observed directly - a measurement will yield one definite value of a physical property even when the state in not an eigenstate of the measured observable (recall Schr\"{o}dinger's famous cat, which is always found to be either dead or alive, but not both). In such cases, nothing in the quantum description dictates the exact result of a measurement. Textbook QM supplements the unitary evolution of Schr\"{o}dinger equation (SE) with a second dynamical law, which spells a non-unitary
break in the evolution upon measurement, a collapse, instantaneously changing the state of the system to an eigenstate of the measured observable. Accordingly, the result will only be determined probabilistically, where the probability is given by the square amplitude of the eigenstate term, a postulate known as the Born rule. This is in stark contrast to classical mechanics, which only exhibits probabilities stemming from ignorance about the exact phase space state of the system, while remaining fully deterministic and local at the fundamental level of the physical laws.\\

While the collapse postulate makes QM perfectly operational, it introduces ambiguity into the theory. Given that any macroscopic object is just an aggregate of microscopic objects, as suggested by the lack of criterion for otherwise distinguishing them, it is not clear why the SE should not suffice for the full dynamical description of any process in nature. And if it were all encompassing, QM would have been deterministic, and not probabilistic, as the collapse and the Born rule maintain. Attempts to give satisfactory explanations to this predicament lead to discussions about the completeness of the quantum description, and different interpretations. The different approaches range from collapse theories such as the \hyphenation{}Ghirardi-Rimini-Weber (GRW) Spontaneous Localization Model  \cite{ECGRW} and successors thereof, through deterministic variable theories such as Bohmian mechanics  \cite{ECBohm}, the objective general-relativistic collapse suggested by Penrose \cite{ECPenrose}, and all the way to the relative state interpretation by Everett \cite{ECEMWI}, which assumes nothing other than the standard axioms. In the latter, the different branches of a superposition are said to represent different co-existing states of reality, where the observation of a certain outcome is attributed to the specific state of the observer that is correlated to it in the superposition. Each of the superposition terms constitutes a ``branching world'', and is part of a Universal deterministically evolving wavefunction. Hence, it is also known as the ``Many Worlds Interpretation'' \cite{ECVMWI}.\\

The novel approach we shall present tries to tackle the difficulties without resorting to the usual notion of collapse. It will inherit the advantages of the MWI without assuming multiple realities. This approach suggests that a complete description of the physical state has to include two state-vectors, forming the ``two-state''. The states evolve independently by the same unitary dynamical law (and same Hamiltonian), but in opposite temporal directions (where the forward direction is defined according to the direction of entropy increase in the observed Universe). Whenever macroscopic objects, i.e. many-particle systems, are entangled with a microscopic system, as in a measurement, the setup/environment selects a preferred basis, while the unknown backward-evolving state selects a definite outcome from the known forward-evolving state, giving rise to a single definite physical reality. Thus, the probabilistic nature of quantum events can be thought of as stemming from our ignorance of the backward-evolving state, reintroducing the classical concept of probability as a measure of knowledge.\\

The decoherence program \cite{ECZurek,ECDeco} has been successful in reducing, locally, the unobserved coherent superposition of macrostates into a mixture of effectively classical states, pointer states. The damping of the interference terms in the pointer states basis is attributed to the near orthogonally of environmental states entangled with them. By tracing out the environmental degrees of freedom, one may unveil the mixed state in which the system and apparatus are given. However, the trace operation is a purely mathematical procedure, which indicates no reduction of the global state to a single definite measurement outcome. Using the backward-evolving state of the TSVF, we shall demonstrate how a selection of a single outcome may be achieved.\\

We will focus our attention on the boundary conditions posed in each realm. In classical mechanics, initial conditions of position and velocity for every particle fully determine the time evolution of the system. Therefore, trying to impose a final condition would either lead to redundancy or inconsistency with the initial conditions. This situation is markedly different in the realm of quantum mechanics. Because of the uncertainty principle, an initial state-vector does not determine, in general, the outcome of a future measurement. However, adding another constraint, namely, the final (backward-evolving) state-vector, results in a more complete description of the quantum system in between these two boundary conditions, that has bearings on the determination of measurement outcomes. The usefulness of the backward-evolving state-vector was demonstrated in the works of Aharonov {\it et al.} 

The emergence of specific macrostates seems non-unitary from a local perspective, and constitutes an effective ``collapse'', a term which will be used here to denote macroscopic amplification of microscopic events, complemented by a reduction via the final state. We will show that a specific final state can be assigned so as to enable macroscopic time-reversal or ``classical robustness under time-reversal'', that is, reconstruction of macroscopic events in a single branch, even though ``collapses'' have occurred. An essential ingredient in understanding the quantum-to-classical transition is the robustness of the macrostates comprising the measuring apparatus, which serves to amplify the microstate of the measured system and communicate it to the observer. The robustness guarantees that the result of the measurement is insensitive to further interactions with the environment. Indeed, microscopic time-reversal within a single branch is an impossible task because evolution was not unitary. Macroscopic time-reversal, which is the one related to our every-day experience, is possible, although non-trivial.\\

A measurement generally yields a new outcome state of the quantum system and the measuring device. This state may be treated as an effective boundary condition for both future, and past events. We suggest it is not the case that a new boundary condition is independently generated at each measurement event by some unclear mechanism. Rather, the final boundary condition of the Universe includes the appropriate final boundary conditions for the measuring devices which would evolve backward in time to select a specific measurement outcome. In the following sections we shall demonstrate how this boundary condition arises at the time of measurement due to a two-time decoherence effect. Indeed, we will see that in the pointer basis (determined by decoherence), the outcome of the measurement can only be the single classical state corresponding to the final boundary condition. We thus suggest a particular final boundary condition for the Universe, in which each classical system (measuring device) has, at the time of measurement, a final boundary condition equal to one of its possible classical states (evolved to the final time). \\

A further requirement is that the final state in the pointer basis will induce, backwards in time, an appropriate distribution of outcomes so as to recover the empirical quantum mechanical probabilities for large ensembles, given by the Born rule. The determination of the measurement statistics by the correspondence between the two states may lead one to conclude that within the framework of TSVF, the Born rule is a coincidental state of affairs rather than a law of nature. That is, that the Born rule is a product of an empirically verifiable, yet contingent, relation between the initial and final boundary conditions, one that needs to be postulated for the sake of deducing the rule. This specific relation is contingent in that it leaves open the possibility of a different relation which will lead to a modified version of the Born rule, while the rest of physics remains where it stands. This, however, is inaccurate. It can be shown that this specific law follows, in the infinite $N$ limit, from the compatibility of quantum mechanics with classical-like properties of macroscopic objects \cite{ECAR,ECBorn}. Under the assumption that for macroscopically large samples, the results of physical experiments are stable against small perturbations, a final state pertaining to the Born rule is the most likely final state, for any ensemble.\\

The work is structured as follows: Sec. \ref{EC2} introduces von Neumann's measurement scheme. It will be used for performing strong (projective), as well as weak measurements throughout the work. The description of quantum reality is then augmented with a final boundary state (Sec. \ref{EC3}) and deterministic operators (Sec. \ref{EC4}). In Sec. \ref{EC5}, averages taken over large ensembles are shown to be deterministic operators representing the cumulative result of a set of weak measurements. Sec. \ref{EC6}, which is the heart of the work, presents the two-time decoherence scheme and the micro-macro quantitative boundary required for attaining time-reversal symmetry. Sec. \ref{EC7} concludes the work.

\section{von Neumann Interaction - Projective and Weak}
\label{EC2}

Throughout this work the von Neumann scheme \cite{ECVN} has a key role in realizing quantum measurements. It was traditionally used for describing projective measurements and the decoherence process \cite{ECVND1,ECVND2,ECVND3}. We shall briefly review these arguments and then discuss an important limiting case where the coupling between the measured system and measuring device is much smaller than the quantum uncertainty of the latter. This case is known as {\it weak measurement}.

Let $S$ denote our system to be measured, prepared at some state $|\psi\rangle$. Suppose $A$ is the
Hermitian operator we wish to measure on the system $S$, having $n$
eigenvectors $|a_{i}\rangle$ such that $A|a_{i}\rangle=a_{i}|a_{i}\rangle$. When  expressed in the
eigenbasis of $A$ the system's wavefunction takes the form:

\begin{equation}
\label{ro}
|\psi\rangle=\sum_{i}\alpha_{i}|x_{i}\rangle.
\end{equation}

Let $|\phi_{d}\rangle$ denote the wavefunction of the measurement
device (also called a ``pointer''). When represented in the position basis it will be written
as:

\begin{equation}
|\phi\rangle=|\phi_{d}\rangle=\int_{q}\phi(q)|q\rangle dq,
\end{equation}

\noindent where $q$ is the position variable of the measuring pointer.
Let $Q_{d}$ be the position operator such that $Q_{d}|q\rangle=q|q\rangle$
(here, we use $Q_d$ to distinguish the operator $Q_d$ from
its eigenvector $|q\rangle$ and eigenvalue $q$, the subscript $d$ is used for denoting the measuring device). It is assumed that initially $\phi(q)$ behaves normally around $0$ with some variance $\sigma^{2}$:

\begin{equation}
\phi(q)=(2\pi\sigma^{2})^{-\frac{1}{4}}e^{-q^{2}/4\sigma^{2}}.
\end{equation}

The measuring device $|\phi_{d}\rangle$ is later examined and the shift in the pointer's position is measured.

\noindent Consider the interaction Hamiltonian $H_{\textrm{int}}$
(\cite{ECTC,ECACE}):

\begin{equation}
H=H_{\textrm{int}}=g(t)A\otimes P_{d}.
\end{equation}

\noindent Here $g(t)$ is a coupling impulse function satisfying:

\begin{equation}
\int_{0}^{T}g(t)dt=g, \label{eq:cif}
\end{equation}

\noindent where $T$ is the coupling time, $g$ is the coupling strength and $P_{d}$ is the
operator conjugate to $Q_{d}$ such that $[Q_{d},P_{d}]=i\hbar$.

We shall start the measurement process with the vector:

\begin{equation}
|\psi\rangle\otimes|\phi(q)\rangle,
\end{equation}

\noindent in the product space of the two systems. Then we apply the
following time evolution based on the interaction Hamiltonian above

\begin{equation}
e^{-i \int H dt/\hbar}|\psi\rangle\otimes|\phi(q)\rangle.
\end{equation}

\noindent It is easy to see that on each of the vectors $|a_{i}\rangle\otimes|\phi(q)\rangle$
the Hamiltonian $H$ takes $Q_{d}$ to $Q_{d}+ga_{i}$,
(Heisenberg evolution):

\begin{equation}
Q_{d}(T)-Q_{d}(0) =\int_{0}^{T}dt\frac{\partial Q_{d}}{\partial t}\nonumber =\int_{0}^{T}\frac{i}{\hbar}[H, Q_{d} ]dt=g a_{i}
\end{equation}

\noindent The corresponding transformation
of the coordinates of the wavefunction is:

\begin{equation}
e^{-i\int H dt/\hbar}|\psi\rangle\otimes|\phi\rangle=\sum_{i}\alpha_{i}|a_{i}\rangle\otimes|\phi(q-ga_{i})\rangle. \label{ECVNent}
\end{equation}

\noindent In case $ga_i>>\sigma$ (that is, the coupling strength is much larger than the pointer's uncertainty), then $|\phi(q-ga_{i})\rangle$ and $|\phi(q-ga_{j})\rangle$ are almost orthogonal for $i \neq j$. Hence, the different possible measurement outcomes are projected into distinct states of the measuring device. By this, the first stage of the measurement (also known as pre-measurement) is over. In experiments, however, we do not observe our measured system entangled with various pointer states - we see only one outcome. Therefore, a second, non-unitary amplification stage is needed. Once amplified due to coupling with the environment (which is considered within this work to be part of the macroscopic measurement device), we would see only one eigenstate out of the initial superposition. This reflects the collapse on a single measurement outcome $a_i$.

Alternatively, when $ga_i<<\sigma$, the functions $|\phi(q-ga_{i})\rangle$ are highly overlapping and hence the measurement result is inconclusive. In this case, known as a weak measurement, the von Neumann interaction does not end in a collapse, nor in an unambiguous information about the measured system. We gather a minute amount of information regarding the system at the price of slightly changing the measured state. When repeating the weak measurements over a large ensemble containing $N$ particles, the outcomes accumulate like independent identically distributed normal variables, hence the relative error drops like $\frac{1}{\sqrt{N}}$.










\subsection{Weak Measurements}

Originally developed to test the prediction of the Two-State-Vector Formalism, Weak measurement \cite{ECAAV} has already been proven to be very helpful in several experimental tasks \cite{EC5,EC5prime,ECWhiteLight,ECJordan}, as well as in revealing fundamental concepts \cite{ECunusual,ECpast,ECRHardy,ECInter,ECCheshire,ECfuture} (which are admittedly are under controversy now). Tasks traditionally believed to be self contradictory by nature such as determining a particle's state \emph{between} two measurements prove to be perfectly possible with the aid of this technique. Within the framework of the TSVF, weak measurements reveal new and sometime puzzling phenomena. For a general discussion on weak measurements see \cite{ECTC,ECACE,EC4f}.

\section{The Final State}
\label{EC3}

We would like to add now an important ingredient to our analysis -  a final boundary state. The idea that a complete description of a quantum system at a given time must take into account two boundary conditions rather than one is known from the two-state vector formalism (TSVF). The TSVF is a time-symmetric formulation of standard quantum mechanics, which posits, in addition to the usual state vector, a second state vector evolving from the future towards the past. This approach has its roots in the works of Aharonov, Bergman and Lebowitz \cite{ECABL}, but it has since been extensively developed \cite{ECReznik,ECTSVF}, and has led to the discovery of numerous peculiar phenomena \cite{ECAR}. \\

The TSVF provides an extremely useful platform for analyzing experiments involving pre- and post-selected ensembles. Post-selection is permitted in quantum mechanics due to the effective indeterminacy of measurement, which entails that the state of a system at one time and its Hamiltonian only partially determine measurement outcomes at later times. {\it Weak measurements} enable us to explore the state of the system at intermediate times without disturbing it. The two-state $\langle \phi |~| \psi \rangle$ created by both boundary conditions allows to define for any operator the {\it weak value}

\begin{equation} \label{ECWV}
\langle A \rangle_w= \frac{\langle \phi |A| \psi \rangle}{\langle \phi | \psi \rangle}.
\end{equation}\\

This weak value naturally appears as the pointer's shift when we perform a weak measurement on a pre-/post-selected ensemble $\langle \psi_f|~|\psi_i \rangle$

\begin{equation}
\begin{array}{lcl}
\langle \psi_f| e^{-i\int H dt/\hbar}|\psi_i\rangle\otimes|\phi\rangle \approx
\langle \psi_f| 1 -ig A\otimes P_{d}|\psi_i\rangle\otimes|\phi\rangle =
\langle \psi_f|\psi_i\rangle \left(1-ig\langle A \rangle_w P_{d}\right)|\phi\rangle \approx \\
\langle \psi_f|\psi_i\rangle e^{-ig\langle A \rangle_w P_{d}}|\phi\rangle =
\langle \psi_f|\psi_i\rangle |\phi(q-g\langle A \rangle_w P_{d})\rangle
\end{array}
\end{equation}

The power to explore the pre- and post-selected system by employing weak measurements motivates a literal reading of the formalism, that is, as more than just a mathematical analytic tool. It motivates a view according to which future and past are equally important in determining the quantum state at intermediate times, and hence equally {\it real}. Accordingly, in order to fully specify a system, one should not only pre-select, but also post-select a certain state using a projective measurement.\\

\section{Deterministic Operators}
\label{EC4}

Describing quantum mechanics through operators within the Heisenberg representation allows to identify {\it deterministic} operators. We find these operators vital for understanding the ``classical-quantum interplay''. Although operating on {\it quantum} states, the outcome is deterministic, as if we have performed a classical experiment with {\it classical} observables. Identifying the set of all deterministic operators with respect to given state, amounts to providing a complete description of the quantum system (equivalent to that of Schr\"{o}dinger) yet somewhat more compact and predictable \cite{ECHeis}.

In the Schr\"{o}dinger representation, a quantum system is fully described by vector in a Hilbert space. Its time evolution is dictated by the Hamiltonian and calculated according to the Schr\"{o}dinger equation. Observables are usually described by time-independent operators. In the Heisenberg representation however, a physical system can be described by a closed set (under addition and multiplication) of {\it deterministic operators}, evolving in time according to the Heisenberg equation, whereas the state does not change in time. Deterministic operators are Hermitian ``eigenoperators'', i.e., Hermitian operators for which the system's state is an eigenstate:

\begin{equation}
\{A_i ~\text{such that}~ A_i|\psi\rangle=a_i|\psi\rangle, a_i \in \Re\}
\end{equation}

It is easy to show that for describing a particle in an $n$-dimensional Hilbert space, a set of $(n-1)^2+1$ deterministic operators, whose eigenvectors span the relevant sub-space, is required \cite{ECToll2009}. The physical significance of these operators stems from the possibility to measure all of them without disturbing the particle, that is, without inducing collapses. Therefore, they can also be measured successively without mutual disturbance

\begin{equation}
[A_i, A_j]|\psi\rangle=0,
\end{equation}

\noindent for any $i,j$. \\


The mathematical equivalence between the Schr\"{o}dinger and Heisenberg representations assures that there is a one-to-one correspondence between the wavefunction and the set of deterministic operators describing the same physical system. However, the wavefunction also expresses non-deterministic properties, such as positions in a delocalized system. We do not consider these to be real properties of the single particle, maintaining that they are exhausted by the set of deterministic properties. If, for instance, the position operator is not deterministic, the question ``where is the particle?'' bares no meaning. The non-deterministic operators do not represent properties intrinsic to the particle. They only reflect probabilistic properties at the ensemble level. It should be noted though, that an ensemble possesses a set of deterministic operators larger than that of the single particle. Importantly, as we shall see in the next section, the average value of any one-particle operator is deterministic. The wavefunction itself is another deterministic operator of an ensemble of particles \cite{ECHeis}. \\

Within a time-symmetric formalism adding a final state amounts to adding a second set of deterministic operators on top of the one dictated by the initial state, thereby enlarging the assortment of system properties. The properties expressed by this two-fold set are the ones which we believe to constitute the primitive ontology of quantum mechanics.

\section{Averages as Deterministic Operators}
\label{EC5}

We claim in this section that most classical experiments we perform are based on weak measurements. In fact, any physical quantity averaged over a large ensemble of particles can be understood as a result of weak coupling to each of them. This follows from the following statement: {\it The state of a large ensemble is nearly an eigenstate of the ``average'' operator} \cite{ECAV}. In this sense, averages are {\it deterministic} operators, thus having a classical nature.

The proof is as follows. We shall utilize the well-known relation

\begin{equation}
A|\Psi\rangle=\bar{A}|\Psi\rangle+\Delta A |\Psi_\perp\rangle,
\end{equation}
where $\Delta A=[(\bar{A^2})-(\bar{A})^2]^{1/2}$ and $|\Psi_\perp\rangle$ is some state orthogonal to $|\Psi\rangle$. For an ensemble of identical particles we thus obtain:
\begin{equation}
\begin{array}{lcl}
\frac{1}{N}\sum_{i=1}^N A_i \prod_{i=1}^N |\Psi\rangle_i = \frac{1}{N}\sum_{i=1}^N ( \bar{A}|\Psi\rangle_i+\Delta A |\Psi_\perp\rangle_i)\prod_{k \neq i}^N|\Psi\rangle_k= \\
= \bar{A} \prod_{i=1}^N |\Psi\rangle_i + \frac{\Delta A}{N}\sum_{i=1}^N|\Psi_\perp\rangle_i \prod_{k \neq i}^N|\Psi\rangle_k,
\end{array}
\end{equation}
where $A_i$ is the observable of interest applied to the i-th particle. Due to mutual orthogonality, the last term's norm is $O(\frac{1}{\sqrt{N}})$ and hence can be neglected. Therefore, the product of $N$ identical states is an eigenstate of any average operator at the limit of $N\rightarrow\infty$. Moreover, the procedure of obtaining it is a weak measurement based on the interaction Hamiltonian $H=g(t)\frac{1}{N}\sum_{i=1}^N A_i$ creating a very weak coupling (scales like the inverse of the ensemble's size) to each particle.

A few generalizations of the above procedure for obtaining the average value can simply follow: \\
(i) If instead of a product of $N$ identical states, we have a slight fluctuations $|\delta\psi\rangle_i$ in each single particle state, then as long as these fluctuations are random, we would be able to neglect them at the limit of large $N$.\\
(ii) If rather than a product of $N$ identical states, we have $N_1$ particles prepared in some state $|\chi\rangle_1$, $N_2$ particles prepared in $|\chi\rangle_2$,...,$N_m$ particles prepared in $|\chi\rangle_m$, such that $\sum_{i=1}^m N_m=N$ and $\forall m~N_m>>1$ the proof can be repeated when again the uncertainty term is negligible. The average would now be $\bar{A}=\frac{\sum_{i=1}^m N_i\bar{A}_i}{N}$, an eigenstate of $A$, achieved through weak measurements. \\
(iii) If instead of a single pointer we have a sequence of $k$ pointers, where $N/k>>1$ we would still be able to utilize the law of large numbers for each of them and arrive at the previous result.\\

Another classical property of the average operators is their almost exact commutativity \cite{ECAR2002}. If, for instance, we have an ensemble of spins prepared in the same state then the average spin along the x-direction is defined by $\bar{S_x}=\frac{\hbar}{2}\frac{\sum_{i=1}^N \sigma_x^i}{N}$, and
\begin{equation}
\lim_{N\rightarrow\infty}[S_x,S_y]= i \hbar \lim_{N\rightarrow\infty} \frac{\bar{S_z}}{N}=0,
\end{equation}
implying commutativity of the average spin operators along the x and y directions (recall the notion of deterministic operators presented in the previous section).

We are the led to the conclusion that our everyday classical experience, which is based on averaged properties of large ensembles, may essentially originate from the theory of weak measurements. Any macroscopic event, perceived by us as deterministic, can be decomposed into a large number of weak measurements, each of which is uncertain.

\section{Time-Reversal Symmetry and the Macroscopic Threshold}
\label{EC6}

We shall see now that a post-selected state of the universe may account for the apparent collapses we see in nature. Furthermore, and strictly related to the above discussion, it will be shown that a time-reversal symmetry sets a lower bound on the number of particles comprising a macroscopic object.

Assume that the world is symmetrically described by an initial, as well as a final, fine-tuned boundary condition. We further assume the existence of a thermodynamic arrow of time set in the direction of entropy increase, hence the final state will characterize a highly entangled world with high entropy. Macroscopic objects (to be quantitatively defined below) are comprised of at least $N>>1$ microscopic elements, that can be coupled to other, external, microscopic elements. As discussed in Sec. \ref{EC2}, quantum measurements will take the form of the von Neumann scheme (assumed in this section to be carried out using a qubit rather than a continuous pointer) and will be followed by a macroscopic amplification (that is, a macroscopic record of the microscopic result encoded in the state of at least $N$ microscopic particles, or ``environment'' in the language of decoherence). As a result of the measurement a ``collapse'' of the measured microscopic degrees of freedom might seem to occur from the experimentalist's point of view. This apparent collapse, resulting from a partial overlap with a specific post-selected state, obeys time-reversal symmetry under several conditions.

To illustrate the suggested scheme we shall discuss first the simplest case where only one measurement is performed and the macroscopic world (including measuring devices) does not collapse. In the next subsection we will analyze the important case in which the distinction between microscopic and macroscopic world depends only on the number of microscopic elements comprising the objects in question. That is, part of the macroscopic measuring device will be assumed to ``collapse'', but nevertheless, our macroscopic world will be shown to maintain its robustness under time-reversal. For further details and generalizations we refer the reader to \cite{ECMP}. \\

\subsection{A single ideal measurement}
\label{sec:4}
Let our system be initially described by
\begin{equation}
\label{system_1}
|\Psi(t_0)\rangle = (\alpha |1 \rangle+ \beta |2 \rangle)|READY \rangle |\epsilon_0 \rangle ,
\end{equation}
where $\alpha|1\rangle+\beta|2\rangle$ is the state of a microscopic particle, $|READY\rangle$ is the pointer state of the measuring device and $|\epsilon_0 \rangle$ is the state of the environment (for simplicity of notation, Normalizations are omitted hereinafter). Following the von Neumann scheme, we create at time $t=t_1$ a coupling between the particle state and the pointer state, establishing a one-to-one correspondence between them. We will denote the orthogonal pointer states by ``I'' and ``II''. The pointer will shift to $|I\rangle$ in case the particle is in $|1\rangle$ and to $|II\rangle$ in case the particle is in $|2\rangle$:
\begin{equation}
\label{coupling_1}
|\Psi(t_1)\rangle = (\alpha |1 \rangle|I \rangle+ \beta |2 \rangle|II \rangle) |\epsilon_0 \rangle.
\end{equation}
Then, in the course of a short time $t_d$, the preferred pointer state is selected and amplified by a multi-particle environment in the process of decoherence. Since the pointer state basis is favored by system-environment interactions, it is not prone to further entanglement and decoherence. Therefore, it enables us to read off the result of the measurement from the environment in which it is encoded in a unitary fashion. The reading of a specific result does not correspond to just one specific state of the apparatus/environment but rather to a subset of states taken from a very large state-space, where distinct readings correspond to orthogonal states. Physically, these may be spatially separated blotches on a photo-detector, or concentration of molecules in a corner of a chamber. We represent these distinct environmental subsets as $\epsilon_1$ and $\epsilon_2$, and the dynamical process is thus
\begin{equation}
\label{decoherence_1}
|\Psi(t_1+t_d)\rangle = \alpha |1 \rangle|I \rangle|\epsilon_1 \rangle+ \beta |2 \rangle|II \rangle|\epsilon_2 \rangle ,
\end{equation}
This is a macroscopic amplification of the microscopic measurement, which results in what we call ``measurement'' of the particle. After this point, the particle may continue to interact with other objects (microscopic or macroscopic).\\
Now comes the crucial part. Let the backward-evolving state at $t=t_f$ contain only a single term out of the preferred pointer basis
\begin{equation}
\label{final_1}
\left\langle \Phi\left(t_f\right)\right|= \langle\phi| \langle I| \langle \epsilon_1|,
\end{equation}
where $\langle\phi|$ is a final state of the microscopic particle. Within the TSVF, our system will be described by the two-state:
\begin{equation}
\label{TSVF_1}
\langle\phi| \langle I| \langle \epsilon_1|~(\alpha |1 \rangle|I \rangle|\epsilon_1 \rangle+ \beta |2 \rangle|II \rangle|\epsilon_2 \rangle),
\end{equation}
for $t_1+t_d<t<t_2$. This is essentially a future choice of $|I\rangle$, which may serve as a reason for the initial outcome represented by the microscopic state $|1\rangle$. The approximate orthogonality of $|\epsilon_1\rangle$ and $|\epsilon_2\rangle$ assures that after reducing the density matrix to include only the observable degrees of freedom, within the interval $t_1+t_d<t<t_2$, only the first term in Eq. \ref{TSVF_1} will contribute, accounting for the macroscopic result we witness. In the most general case, the backward environment-pointer state need not be exactly identical to the corresponding term in the forward-state, as long as the measure of its projection on it is exponentially (in the number of particles) larger than the measure of its projection on the non-corresponding term(s) of the forward state.\\

Regarding weak values (Eq. \ref{ECWV}), if any were measured during the intermediate times, they would have been determined by the specific selection made by the final state. Moreover, any interaction with this pre- and post-selected ensemble would reflect this final state in the form of the ``weak potential'' \cite{ECunusual}.\\

The effective boundary condition for the past of the backward-evolving state determines the observed measurement outcome by a backward decoherence process. That is, just the same as the backward-state sets the boundary for the future of the forward-evolving state, the forward-evolving state sets the boundary for the past of the backward-state. Together with the regular decoherence, this amounts to a symmetric two-time decoherence process \cite{ECGruss1,ECGruss2}, allowing for a generalization to multiple-time measurements. This subject is formalized in the next subsection, where we present a detailed description of two consecutive measurements. \\

The important conclusion we should bear in mind is related to the time-reversed process. Starting from the state $|\Psi(t_f)\rangle$ as described by Eq. \ref{final_1} and going backwards in time, we are able to reconstruct the pointer reading $|I\rangle$ although the measured microscopic particle has changed its state. This relates to the concept of ``macroscopic robustness under time-reversal'' on which we elaborate within the next subsection. \\

We note that in cases where the free Hamiltonian is non-zero, we will have to apply the forward time-evolution operator on the final boundary condition, which would then cancel upon backward time evolution to the present state. This clearly does not change the results, and therefore we preferred to discuss a zero Hamiltonian.\\

\subsection{When macroscopic objects also collapse}
\label{sec:6}

We are now in position to address the issue of time reversibility of the dynamical equations governing the macrostates. While this property is most naturally present in the MWI, as long as the macroscopic objects stay intact, it may also exist in a single branch, even if its history includes non-unitary events. This is due to the fact that the result of a measurement performed on a microscopic state is stably stored within the macroscopic objects, as we have seen in the last chapter, and can theoretically be extracted. Therefore, while the measured microstate may change non-unitarily from our local perspective, our measurement reading may not. This possibility is what we will refer to as ``classical robustness under time-reversal''. As well as being a landmark of classical physics, time reversibility is vital in order to draw valid conclusions about the early Universe from our current observations.\\

Let the system not be completely isolated, and allow external quantum disturbances which interfere with the evolution in an indeterministic and thus irreversible way (from the single-branch perspective). It is generally accepted that the pointer states selected by the environment are immune to decoherence, and are naturally stable \cite{ECZurek,ECDeco}. Problems start when the (macroscopic) measurement devices begin to disintegrate to their microscopic constituents, which may couple to other macroscopic objects and effectively ``collapse". These collapses seem fatal from the time-reversal perspective, as time-reversed evolution would obviously give rise to initial states very different from the original one. To tackle this, we demand that subtle environmental interactions, mildly altering the macrostate, will not stray too far from the subset of states indicating the perceived measurement result, compared to the orthogonal result.\\

Considering the free evolution of the measuring device and applying it backwards from the final and slightly altered state, the state at the time of measurement will still project heavily onto the same sub-space, indicating the same reading. This may be regarded as macroscopic physics having time-symmetric dynamics. While it might be the case that we do not reconstruct the starting microscopic configuration, being macroscopic objects, this should not upset us as long as our experience remains the same, that is, as long as macroscopic readings, depending on the macrostate of their $N$ micro-particles, do not change when backward evolution is applied. This will be shown to be the case when several assumptions are made regarding the macroscopic objects and the rate of collapse.\\

To derive the limit between microscopic and macroscopic regimes we will assume now that the amplification mechanism consists of at least $N>>1$ particles belonging to the environment or measuring device, from which only $n<<N$ particles may later be measured and collapsed without rendering the dynamics irreversible. by ``measured" we do not necessarily mean that an observer entangled them with a device designated for measurement. Rather, we mean that they may get entangled with some other degree of freedom and decohere. We believe that it is reasonable to assume that $n<<N$ always, because measuring $N$ (which is typically, $10^{23}$) particles and recording their state is practically impossible. Eqs. \ref{system_1}-\ref{decoherence_1} still have the same form, but the measurement of the environment at some $t=t_2$ leads to
\begin{equation}
\label{particle_2}
|\epsilon_i(N)\rangle~~\rightarrow~~\prod_{j=1}^{N-n}|C_i^{(j)}\rangle|\epsilon_i(N-n)\rangle
\end{equation}
for $i=1,2$ representing encodings of two orthogonal microstates. The environment states are $N$-particles states at first, and later contain only $N-n$ particles, while their other $n$ components ``collapse'', for simplicity of calculation, to the product state $\prod_{j=1}^{n}|C_i^{(j)}\rangle$.
The trivial point, although essential, is that
\begin{equation}
\langle C_1^{(j)}|\epsilon_1^{(j)}\rangle =\gamma_1^{(j)} \neq 0,
\end{equation}
for every $j=1,2,...,n$, where $\epsilon_1^{(j)}$ is the j-th environment state before the collapse, i.e. collapse can never reach an orthogonal state. For later purposes let us also assume
\begin{equation}
\langle C_2^{(j)}|\epsilon_1^{(j)}\rangle =\gamma_2^{(j)} \neq 0
\end{equation}
It is not necessarily different from $0$, but as will be demonstrated below, this is the more interesting case. We would like to show that the final boundary state of Eq. \ref{final_1} still has much higher probability to meet $|\epsilon_1\rangle$ than $|\epsilon_2\rangle$, and hence the pointer reading is determined again by the specific boundary condition, despite the collapse of the classical apparatus. Indeed, under the assumption of ending the evolution in the following final boundary condition:
\begin{equation}
\label{final_2}
\left\langle \Phi\left(t_f\right)\right|= \langle\phi| \langle I| \langle \epsilon_1|,
\end{equation}
we can define the ``robustness ratio'' as a ratio of probabilities: The probability to reach backwards in time the ``right'' state $|I\rangle$ divided by the probability to reach the (``wrong'') $|II\rangle$ state. This ratio ranges from zero to infinity suggesting low (values smaller than $1$) or high agreement (values grater than $1$) with our classical experience in retrospect. In our case it is
\begin{equation}
\label{RR}
\frac{Pr(Right)}{Pr(Wrong)}=\frac{\prod_{j=1}^n\gamma_1^{(j)}}{|\langle \epsilon_1(N-n)| \epsilon_2(N-n)\rangle|^2\prod_{j=1}^n\gamma_2^{(j)}}\simeq |\langle \epsilon_1(N-n)| \epsilon_2(N-n)\rangle|^{-2}
\end{equation}
Hence for a sufficiently large ratio of $N/n$ ``classical robustness'' is attained - the result of Eq. \ref{RR} is exponentially high. If one is not ready yet to accept that below some minimal probabilistic threshold events do not occur in our universe (and hence, only the outcome chosen by the boundary state occurs) we can choose a better tuned boundary condition allowing this ratio to diverge (i.e. at least one of the $\gamma_2^{(j)}$ factors is zero).

The significance of the above result is the following: even though from the perspective of the single branch a non-unitary evolution has occurred, there exists a final boundary state which can reproduce with high certainty the desired macroscopic reality when evolved backwards in time. This ``robustness ratio'' can be used also for the definition of macroscopic objects, i.e. defining the border between classical and quantum regimes. We thus understand the significance of macroscopic objects in storing information for extended times until the final boundary condition arrives and information about past events can be released \\

We see that at the cost of having to introduce a final boundary condition, we are able to reclaim determinism (in the two-state sense) and ensure macroscopic time-reversal. It was already assumed, that subsequent to the measurement interaction, decoherence causes an effectively irreversible branching of the superposition into isolated terms. Therefore, no inconsistencies can arise from the existence of a special final boundary condition of the form described before, which simply causes the selection of a single specific branch from the many worlds picture. In this view, the measurement process does not increase the measure of irreversibility beyond that of regular thermodynamics. Additionally, accounting for the apparent collapse, TSVF does not suggest a microscopic quantum mechanical arrow of time. It does however assume asymmetric initial and final boundary conditions.\\

According to the TSVF, any post-selected state which is not orthogonal to the pre-selected state is permissible. However, in our model we have discerned a special boundary condition which accounts for the experimental result, i.e. for the single outcomes which actually occurred in measurements that were actually performed, as well as for the Born rule statistics. This choice is justified on several grounds:\\
(i) It unites the two dynamical processes of textbook quantum mechanics - the SE and collapse - under one heading. In doing so, it renders QM deterministic and local on a global level, and above all, rids of the ambiguity involved in the approach of ``unitary time evolution+non-unitary collapse''. \\
(ii) It allows for robustness under time-reversal of macroscopically large systems.\\
(iii) It is a natural framework to understand weak values and weak reality.\\
(iv) As explained, it is not the case that we could have chosen any sort of boundary condition and still maintain classicality on the macro-level. Only states pertaining to the Born rule allow for that. So empirical observations other than the Born rule by itself (e.g. stability under perturbations) can be seen as supporting evidence for a backwards-evolving state with these properties, if any at all.\\
We find these reason enough to postulate such a boundary condition. Moreover, the initial state of our universe can also be regarded as ``unique'', and therefore, we would like to perceive these two boundary conditions as reasonable, constructive and even necessary for explaining our current observations, rather than artificial. It should also be stressed that in spite of this ``uniqueness'', the final state has high thermodynamical entropy and also high entanglement entropy, since it encodes all the measurement outcomes of microscopic objects.\\

It should be noted that $n$ cannot grow to be $N$, i.e., there is always a ``macroscopic core'' to every macroscopic object which contained initially $N$ or more particles. It can be shown that $\frac{dn}{dt} \le 0$ and also that $\frac{dn}{dt} \rightarrow 0$ for long enough times, assuming for example an exponential decay of the form:
\begin{equation}
N(t)=N(0)exp(-t/T),
\end{equation}
where $T$ is some constant determining the life time of macroscopic objects. Also, on a cosmological scale (inflation of the universe) it can be shown that after long time, measurements become less and less frequent (macroscopic objects which can perform measurements are simply no longer available). That means there is more than one me

The MWI was invoked in order to eliminate the apparent collapse from the unitary description of QM. Within the MWI, the dynamics of the universe is both symmetric and unitary. We have now shown that these valuable properties can be attained even at the level of a single branch, that is, without the need of many worlds, when discussing macroscopic objects under suitable boundary conditions. Despite the seemingly non-unitary evolution of microscopic particles at the single branch, macroscopic events can be restored from the final boundary condition backwards in time due to the encoding of their many degrees of freedom in the final state.

\section{Conclusions}
\label{EC7}
Three complementary approaches for understanding the transition from quantum to classical physics were discussed: deterministic operators, weak measurements and post-selection.

We began with the observation that deterministic operators capture in some sense the gist of classical determinism even when applied to quantum systems. We then showed that in quite general cases, the quantum average is such a deterministic operator. As an offshoot of this approach, we have seen that a measurement of a classical average is in fact a quantum weak measurement of each particle in the ensemble. Finally, the addition of post-selection and the requirement for classical robustness under time-reversal then helped us defining the border line between classical and quantum regimes.

We hope that in future works we will be able to strengthen the relations between the three approaches and combine them into a coherent, fresh perspective on the quantum ontology.

\begin{acknowledgments}

We would like to thank Tomer Landsberger for many helpful discussions. E.C. was supported by ERC AdG NLST. Y.A. acknowledges support from Israel Science Foundation Grant No. 1311/14, ICORE Excellence Center \textquotedblleft Circle of Light\textquotedblright and the German-Israeli Project Cooperation (DIP).

\end{acknowledgments}

\end{document}